# On the applicability of distributed ledger architectures to peer-to-peer energy trading framework

Van Hoa Nguyen, Yvon Besanger
University Grenoble Alpes, Grenoble INP,
CNRS, G2Elab,
F-38000 Grenoble, France
van-hoa.nguyen@grenoble-inp.fr;
yvon.besanger@g2elab.grenoble-inp.fr

Quoc Tuan Tran, Minh Tri Le
CEA-INES
50 Avenue du Lac Léman,
73370 Le Bourget-du-lac, France
quoctuan.tran@cea.fr
minhtri.le@cea.fr

*Abstract* - As more and more distributed renewable energy resources are integrated to the grid, the traditional consumers have become the prosumers who can sell back their surplus energy to the others who are in energy shortage. This peer-to-peer (P2P) energy transaction framework benefits the end users, financially and in term of energy security; and the network operators, in term of flexibility in DRES management, peak load shifting and regulation of voltage/frequency. Environmentally, P2P energy transaction also helps to reduce carbon footprint, reduces DRES payback period and incentivizes the installation of DRES. The current centralized market model is no longer suitable and it is therefore necessary to develop an adapted decentralized architecture for the advanced P2P energy transaction framework intra/inter-microgrid. In this paper, we discuss several distributed ledger approaches for such framework: Blockchain, Block Lattice and Directed Acyclic Graph (the Tangle). The technical advantages of these architectures as well as the persistent challenges are then considered.

## I. Introduction

The European energy roadmaps[1] and implementation plans [2] require a stronger linkage between power technology research and energy market development. Microgrid and the framework of energy transaction intra/inter-microgrid are considered among the major elements to achieve the desired decarbonized scenario. While the definitions in the literature may vary; in general; a microgrid is attributed with two main properties: firstly, the microgrid has clear electrical boundaries and can act as a single controllable entity with respect to the marcogrid; secondly, microgrid can operate in either grid-connected or islanded modes. To enable the independent functionality, a microgrid may require a number of distributed energy resources (DER) and may have its own centralized or distributed load control systems. On the one hand, microgrid concept eases the challenge of controlling large number of DER and management of peak demand for grid operators. It allows better penetration of renewable electricity generation and improves reliability of supply. On the other hand, the end users benefit a more self-sufficient, autonomous energy provision with less reliance on the electrical grid, as well as resilience in disaster relief. In term of environmental impact, as generations shift towards renewable energy and loads grow locally, efficiency of generation and transmission is increased and microgrid development can contribute significantly to reduce carbon footprint.

Distributed renewable energy resources (DRES) allow consumers to become prosumers as they can feed back the generated energy to the grid. Due to the intermittent nature, microgrid or individual users relying on renewable energy will have to deal with energy surplus or deficit in particular scenarios. A peer-to-peer (P2P) energy transaction framework provides prosumers with the possibility to trade their over-generated energy to others who are in need, or fulfill their consumption with complementary energy supplied from elsewhere. Such a trading framework benefits both the network operators (in term of peak load management, regulation of voltage/frequency, etc.) and the end users (in term of energy security and reducing carbon footprint) [3]. Additionally, by shortening the payback period of DRES installation, the peer-to-peer energy trading framework helps to incentivize the implantation and trading of renewable energy. Energy trading can be of different scale, either among individual prosumers intra-microgrid, inter-microgrid or among DSO, oriented towards different purposes (i.e. power balancing, voltage/frequency regulation in large scale or economic influences in small scale). In [4], the authors suggested a classification of existing P2P projects to three levels according to the hierarchical nature of the distribution networks: P2P within a microgrid, P2P within a CELL (multi-microgrids) and P2P among CELLs (multi-CELLs). In general, it is commented that the existing projects designed business models and little effort was put to the relevant ICT and control systems for such markets, thus energy transactions were considered from the microgrid level and not in intra-microgrid level.

On the other hand, more and more use-cases of micro-energy trading are being investigated, e.g. energy trading

The participation of G2Elab and CEA Ines in this research is partly supported by the French Carnot Institute "Energies du Futur" (http://www.energiesdufutur.eu/), in the PPInterop 2 project and the European Community's Horizon 2020 Program (H2020/2014-2020) under project "ERIGrid" (Grant Agreement No. 654113, www.erigrid.eu).

among plug-in hybrid electric vehicles (EV) [5] in peak load shifting scenario. It is bringing in microgrid technologies and energy market together to present advanced solutions for interaction and exchange of electricity from bottom-up manners. It leads to the necessity of a more secured P2P decentralized electricity trading system with solid market model that can preserve the privacy of the participants[6], in which the new Blockchain technology is considered by many initiatives as a potentially suitable solution.

In this paper, we review the advantages and drawbacks of Blockchain approach for P2P energy trading framework and also point out possible alternative technologies (i.e. Block Lattice and Directed Acyclic Graph). Potential implementation challenges and possible solutions are then discussed. The paper is organized as following: in section II, the emerging tendency of applying Blockchain technology to P2P energy trading is discussed. While there are quite a good number of Blockchain initiatives in this domain, we point out in section III important challenges that may limit or eventually prohibit the application of Blockchain in P2P energy transaction. In section IV, potential alternative distributed frameworks such as Block Lattice and Directed Acyclic Graph are also investigated. In section V, we discuss the common issues of such frameworks and possible solutions; before concluding the paper in section VI.

## II. P2P ENERGY TRANSACTION AND THE EMERGING APPLICATIONS OF BLOCKCHAIN TECHNOLOGY

Nowadays, P2P energy transaction is done most of the time via intermediate of a network operator, using the marcogrid. In this "pseudo" P2P model, the retailer can act as a trusted party to both peers, that can confirm and enable the cooperation. In some case, the individual prosumers are required to directly sell their generated energy to the grid, which requires an independent meter. The grid will then deliver energy to the consumers in separated transactions. The model is in fact centralized and is vulnerable to problems such as single point of failure and privacy leakage [6]. Moreover, the existence of this "middle man" also introduces additional transaction fees or unbalanced buy/sell prices. The development of a fully decentralized P2P energy trading system suffers from several challenges: Trust issues among participants, unbalance between supply and demand and lack of supporting infrastructures.

Recently, several worldwide initiatives have explored the application of Blockchain technology[7] to P2P energy trading. Blockchain technology, at a high level, employs a decentralized ledger of all transactions across a peer-to-peer network and is validated by peers, instead of a single centralized authority. The principle of Blockchain is combining information (e.g. transactions, records or items) in a block of a predefined length and connecting the new blocks to the last block in the chain using hash functions (Figure 1). The participants have to sign their transactions inside the block with their private key. In this "hash-based" architecture, it is not possible to modify or delete present data in the system. As a result, the information is transparent and it is possible to determine who issued the information in the block as well as to track the sequence of changes. The immutability of information in therefore ensured.

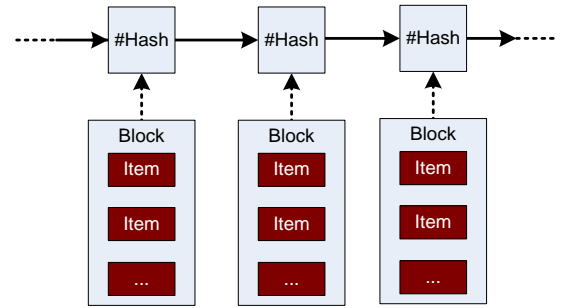

Figure 1: Concept of Blockchain

In general, this decentralized architecture, while removes the need of an arbiter, still ensures the verifiability of the participants (organization, individuals, devices) thank to the permanent record of exchanges and information from devices stored on the Blockchain, thus, security and transparency is enhanced while transaction time and fee is reduced. The principal difference of P2P energy trading framework using Blockchain with traditional framework is that all participants can execute energy exchanges with any other members without the restriction of a centralized authority. It enables the possibility for the consumer to negotiate directly with the prosumers and determine the value (which could even be 0) via smart contracts and to choose the type of energy to consume independently of the marcogrid. Moreover, it provides a bidding system for preferred energy type in case of unbalance of supply/demand of the desired type of energy (e.g. solar).

The application of Blockchain in energy transaction has been developed in parallel with the evolution of Blockchain technology (i.e. Blockchain 1.0, 2.0 and 3.0) [8]. In energy section, first application of Blockchain 1.0 is shifting the billing system towards cryptocurrencies and is extended to enabling physical transaction using smart contracts (2.0). Some initiatives of using Blockchain 1.0 in energy transaction can be named: Solarcoin[1], Share&Charge[2]. Using Blockchain 2.0 as a platform for more complex services and interactions, we can mention some energy trading projects such as: EXERGY (USA)[3], Electron (UK)[4], Sunchain (France)[5], Power Ledger (Australia)[6] and Grid Singularity (Austria)[7] etc. Blockchain 3.0, aiming towards big data deployment, is promising for data assessment in smart grid; however, the implementation of the technology has not yet been reached.

---

[1] https://solarcoin.org
[2] https://shareandcharge.com/
[3] https://lo3energy.com/
[4] http://electron.org.uk/
[5] http://sunchain.fr/
[6] https://powerledger.io/
[7] http://gridsingularity.com

## III. CHALLENGES IN BLOCKCHAIN TECHNOLOGY APPLICATION TO A P2P ENERGY TRADING FRAMEWORK

While many worldwide projects are being carried out on the application of Blockchain technology in energy system, both in term of processes and platforms, Blockchain deployment still suffers from various criticism on the technology itself as well as great challenges on its potential applications to energy system.

The technology is strongly criticized over its intensive energy consumption. As estimated by the Digiconomist real-time tool[8], by January 2018, a single Bitcoin (the most popular cryptocurrency based on Blockchain) transaction consumes 385 kWh of electricity, equivalent to 188.45 kg of $CO_2$ emission. This huge energy consumption is explained by the increasingly intensive calculation cost required when the system scales up [9]. This high calculation load potentially prevents the application of Blockchain to an automated P2P energy trading framework (e.g machine-to-machine transaction), due to the hardware requirement to fulfill such requirements.

While there are several ways to improve this energy index, such as using private consortium Blockchain or replacing proof of work with proof of stake; it is still hard to reach a sustainable carbon footprint due to the decentralized architecture that requires a lot of communication among peers.

Naturally, the question is that whether or not the promising interests of applying Blockchain technology to P2P energy trading (incentivizing DRES installation, peak load shifting and power balancing) can justify the high carbon footprint required for its functionality.

Intensive calculation load also leads to the second question of scalability and performance of the technology in large scale system. An energy trading platform requires much stricter constraints in term of security and reactivity, than an ordinary trading platform. The concept of transaction fee for transactions of any value is also a drawback of Blockchain technology as it does not incentivize micro-transactions[10] which is, on the contraire, one of the important feature of a P2P energy trading platform. Therefore, it is debatable if Blockchain technology is suitable for such framework and adaptation of the technology upon implementation is expected.

Another potential issue is the lack of (digital and electrical) platforms to facilitate interaction between peers, either in a P2P network of prosumers in a microgrid or an inter-microgrid market. A Blockchain based energy market has to go beyond financial transaction in cryptocurrencies and has to reflect the physical configuration of the power grids. Furthermore, applications of Blockchain within current regulatory system of energy trading may be a potential problem. As emphasized in [7], in most countries, tax and grid fees contribute a great part of the current energy tariffs. Their implication in P2P energy trading with Blockchain is still missing and can be a potential problem for further development. While the concept of transaction fee in Blockchain can be adapted as a solution to reflect this cost, it is expected to require a lot of work in term of implementation and making regulations. In any case, it is unlikely that the overall financial benefit (for prosumers and for network operators) would be significant enough to incentivize the adoption and the participation into the framework.

In general, while presenting several potential advantages, Blockchain technology still has drawbacks and challenges in term of its applications to P2P energy trading framework. We discuss in the next section, the potential alternative distributed ledger technologies that can be applied as a model for a P2P energy trading framework. Same as the case with Blockchain technology, the advantages and drawbacks of these technologies in a P2P energy trading context will be also discussed.

## IV. POTENTIAL ALTERNATIVE TECHNOLOGIES

Prominently, Block Lattice and Directed Acyclic Graph (DAG – or the Tangle) are considered as improved decentralized platforms for P2P trading over Blockchain. We do not consider the associated cryptocurrencies, but are rather interested in the underlying architectures and transaction mechanisms of these technologies and the possibility of their adoption to a P2P energy trading context.

### A. Block-lattice

Block Lattice alters the idea of Blockchain and introduces a simple transaction mechanism. In general, each participant possesses one private Blockchain which is replicated to all peers in the network (i.e. block lattice) (**Figure 2**). The individual Blockchain is controlled by a private key, thus only the owner can add blocks to it.

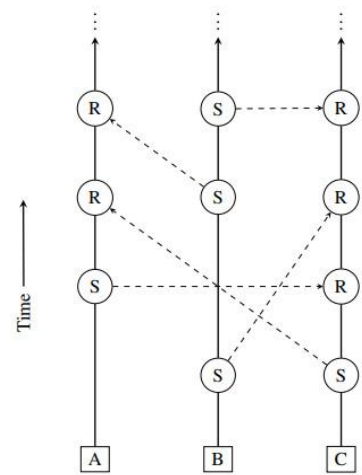

Figure 2: Concept of block-lattice [11].

P2P transaction is performed through the asynchronous "send" (reducing the balance of account) and "receive" (increasing the balance of the account) blocks in the private

---

[8] https://digiconomist.net/bitcoin-energy-consumption

Blockchain of the two peers. The sender creates two blocks "send" and "receive" blocks and signs them with its private key. The "send" block is then added to the sender's Blockchain and the "receive" is sent to the receiver, which will need to also sign the "receive" block with its private key before adding to its Blockchain.

In case of vulnerability and/or conflict, Block Lattice uses a weighted voting system with "representatives" which is basically a trustworthy (wealthy) address that acts as the arbiter [11].

*B. Directed Acyclic Graph – The Tangle*

The Tangle, on the other hand, is based on the idea of a "blockless" distributed ledger [10]. Instead of maintaining a global Blockchain, the DAG requires the node which issues the transaction to perform some computation on two previous transactions in the network. Once the proof of work is finalized, the bundle of data is broadcasted over the network and will be confirmed by future transactions. Transactions are attached to a Directed Acyclic Graph – called the Tangle (**Figure 3**). A weight is added to every transaction and each transaction accumulates weight from the coming transactions that directly or indirectly approve it. When a transaction is successfully confirmed by a certain numbers of other transactions (gains enough weight), it will be attributed a "confirmed" status.

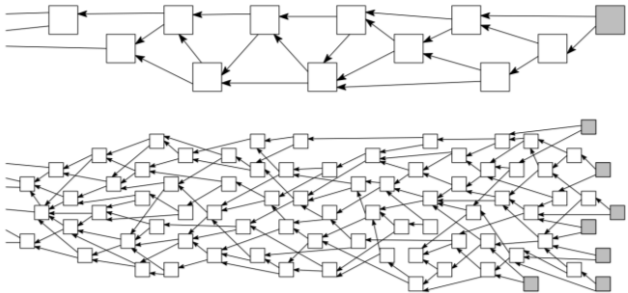

Figure 3: Illustration of DAG in scenarios of low load and high load of incoming transaction flow[10].

This architecture solves the scalability problem of the original Blockchain technology, where it is difficult to achieve consensus when the network grows large. In [12], the authors proved the existence of Nash equilibrium for DAG.

Generally, both Tangle and Block Lattice provide good supports for micro-transactions, which is essential for an intra-microgrid P2P energy network. Both technologies address the limitation of current Blockchain and can potentially be used as the base for P2P energy trading. Several use-cases have already been investigated, e.g. P2P energy trading among PHEV, private transactions and smart contracts. However, while technically improved over Blockchain (e.g. more scalability, feeless and reduced carbon footprint), several challenges still persist in establishing a P2P energy trading framework based on these technologies, as we will discuss in next section.

V. DISCUSSION

We discuss in this section the possible scenarios of application of the aforementioned technologies as architecture for a P2P energy trading framework. Providing better support for micro-transaction as well as requiring less energy for functionality, Block Lattice and DAG can be considered as more suitable compared to Blockchain. However, regardless of the technology, several challenges and issues persist at the level of electrical support and of regulation.

The first challenge is the dependency on the marcogrid as platform provider for energy transaction. While energy transaction intra-microgrid can be done via individual support (even though it is not always the case nowadays), it is expected that the energy transaction inter-microgrid still needs to use the marcogrid. Eventually, in the larger context load shifting or power balancing, we can hardly separate the intra/inter-microgrid P2P energy trading framework from marcogrid and achieve a total P2P model. It is therefore necessary to take into account the fee for utilization of the infrastructure in every transaction. This aspect is however not covered by any of the aforementioned architectures and adaptation is needed before implementation. A possible solution can be suggested: the network operator will no longer act only as a power supplier but also as an infrastructure provider for energy transaction among the sellers and the one in need (i.e. infrastructure as a service) and the fee is paid by the prosumer. This fee can eventually be modulated to include the implication of tax in the second point. The difference of this architecture with current model is that the price is no longer decided by the network operator, but by the bidding system and smart contracts.

The second point that needs to be considered is the lack of adapted judicial guidelines due to the novelty of the framework: in term of regulation of functionality of P2P energy trading and in term of implication of tax and grid fee in transaction. It is even more necessary when the interested architecture is based on distributed ledger due to the anonymity of the individuals/organizations behind the node, even if the transaction record of such node is transparent. It is expected that the technology, upon implementation in reality, will need to be adapted to follow the local regulation (e.g. peers will need to disclose their identities upon registration to the framework).

Environmental challenge is considered as the most important obstacle preventing the implementation of such distributed ledgers technologies. High carbon footprint and energy consumption is expected to be necessary for the functionality of the framework. As estimated in [13], every GB of data requires about 5kWh to support and only 38% of those costs are paid by the end user. Intensive calculation load aside, due to the distributed architecture of Blockchain/Block Lattice, while it seems that the individual does not require a lot of exchange to complete a transaction; the data exchange of the system as a whole is huge. As a result, the carbon footprint when the system scales up is certainly going to be problematic. Among the considered technologies, DAG appears to require much less communication. In any case, consideration of environmental impact is strongly recommended.

On another point of view, there models are based on non-government-backed cryptocurrencies with high volatility and ambiguous notion of value. While it accelerates and secures the transactions, as well as provides fast and feeless transnational trading; employing a non-government-backed token as currency for the framework is susceptible to be problematic and is not convincing for prosumers as well as infrastructure providers. Upon implementation of any of these architectures to P2P energy trading framework, it is necessary to establish links with traditional financial system (e.g. replacing token with government-backed cryptocurrencies or establishing a reliable conversion between token and government-backed currency).

Generally, Blockchain, Block Lattice and DAG introduce potential architectures and innovative transaction mechanisms to P2P energy trading framework, applied to both levels: intra/inter-microgrid. However, there are still a lot of challenges that need to be solved before implementing of such architectures: high carbon footprint, lack of regulation. Most importantly, these technologies can only solved the problem at the information layer and the construction of such P2P energy trading framework still depend fundamentally on the adaptation of electrical layer.

It is also necessary to consider that the main purpose of the framework is matching people with surplus energy to the ones in need, according to several specifications and constraints (e.g. load shifting or requirement of renewable energy). While the discussed approaches excel in term of transparency, verifiability and ease of smart contract creation, it is necessary to explore also other simpler options such as the classical multi-agent approach.

## VI. Conclusion

In conclusion, with high penetration of DRES in microgrids and strong digitalization, P2P energy trading among prosumers intra-microgrid and inter-microgrid will be beneficial to both the network operators and the end users, technically, financially and environmentally. In order to establish such a framework, it is necessary to shift from centralized to decentralized market model.

Blockchain is currently an approach that is adopted by various worldwide P2P energy trading initiatives due to its many advantages. It was discussed in this paper that Blockchain technology still suffers from several drawbacks: intensive calculation requirement and high carbon footprint, lack of scalability and adaptation for micro-transaction. Block Lattice and DAG are interesting alternatives approaches that provide more scalability and are expected to have reduced carbon footprint. However, some challenges persist in term of lack of adaptable platform for P2P energy transaction as well as an adapted regulatory framework. In parallel with establishing the market model and technical support, it is necessary to solve these challenges in the way towards a sustainable P2P energy trading framework.